\documentclass[rnote]{aa}
\usepackage{epsfig}          
\usepackage{graphicx}        
\usepackage{amssymb}         
\usepackage{color}           
\usepackage{url}             
\usepackage{rotating}
\usepackage{enumerate}
\usepackage{tabularx}
\usepackage{setspace}
\usepackage[pdfborder={0 0 0 },urlcolor=blue,breaklinks]{hyperref}
\usepackage[varg]{txfonts}
\usepackage{color}

\begin{document}

\title{Comparison of 30 THz impulsive burst time development to microwaves, H$\alpha$, EUV, and GOES soft X-rays}

\author{R.~Miteva\inst{\ref{inst1},\ref{inst2}}\and
P.~Kaufmann\inst{\ref{inst1},\ref{inst3}}\and
D.~P.~Cabezas\inst{\ref{inst1}}\and
M.~M.~Cassiano\inst{\ref{inst1}}\and
L.~O.~T.~Fernandes\inst{\ref{inst1}}\and
S.~L.~Freeland\inst{\ref{inst4}}\and
M.~Karlick\'{y}\inst{\ref{inst5}}\and
A.~Kerdraon\inst{\ref{inst6}}\and
A.~S.~Kudaka\inst{\ref{inst1}}\and
M.~L.~Luoni\inst{\ref{inst7}}\and
R.~Marcon\inst{\ref{inst8},\ref{inst9}}\and
J.-P.~Raulin\inst{\ref{inst1}}\and
G.~Trottet\inst{\ref{inst6}}\and
S.~M.~White\inst{\ref{inst10}}
}

\institute{
Center of Radio Astronomy and Astrophysics, Engineering School, Mackenzie Presbyterian University, S\~{a}o Paulo, SP, Brazil\label{inst1}
\and
Space Research and Technology Institute, Bulgarian Academy of Sciences, Sofia, Bulgaria \email{rmiteva@space.bas.bg}\label{inst2}
\and
Center of Semiconductor Components, State University of Campinas, Campinas, SP, Brazil\label{inst3}
\and
Lockheed Martin Solar and Astrophysics Laboratory, Palo Alto, CA 94304, USA\label{inst4}
\and
Astronomical Institute of the Academy of Sciences of the Czech Republic, Ond\v{r}ejov, Czech Republic\label{inst5}
\and
LESIA-Observatoire de Paris, Meudon, France\label{inst6}
\and
Instituto de Astronom\'{i}a y F\'{i}sica del Espacio (IAFE), CONICET-UBA, Buenos Aires, Argentina
\label{inst7}
\and
`Gleb Wataghin' Physics Institute, State University of Campinas, Campinas, SP, Brazil\label{inst8}
\and
`Bernard Lyot' Solar Observatory, Campinas, SP, Brazil\label{inst9}
\and
Air Force Research Laboratories, Space Vehicles Directorate, Albuquerque, NM 87117, USA\label{inst10}
}

\date{Received <date>/ Accepted <date>}

\abstract
{
The recent discovery of impulsive solar burst emission in the 30 THz band is raising new interpretation challenges. One event associated with a GOES M2 class flare has been observed simultaneously in microwaves, H$\alpha$, EUV, and soft X-ray bands. Although these new observations confirm some features found in the two prior known events, they exhibit time profile structure discrepancies between 30 THz, microwaves, and hard X-rays (as inferred from the Neupert effect). These results suggest a more complex relationship between 30 THz emission and radiation produced at other wavelength ranges. The multiple frequency emissions in the impulsive phase are likely to be produced at a common flaring site lower in the chromosphere. The 30 THz burst emission may be either part of a nonthermal radiation mechanism or due to the rapid thermal response to a beam of high-energy particles bombarding the dense solar atmosphere.}

\keywords{Sun: flares - Sun: chromosphere - Sun: radio radiation - Sun: UV radiation}

\titlerunning{30 THz impulsive burst on 1 August 2014}
\authorrunning{Miteva et al.}
\maketitle

\section{Introduction}

The terahertz (THz) band is the last unexplored part of the electromagnetic spectrum in solar flare observations. Earlier observations addressed emission from the quiet Sun and quiescent active regions \citep{1970SoPh...14..112T,1981ApJ...247..348L,1999ASPC..183..559G,2008PASP..120...16M,2010SoPh..264...71C}. Weak and rapid mid-IR pulsations, associated with GOES B and C-class flares \citep
{2006PASP..118.1558M,2009PASP..121.1296M,2010SoPh..264...71C}, were detected. 

The first intense impulsive 30 THz bursts were recently reported. One burst reported on 13 March 2012 was associated with a GOES class M8 soft X-ray event. This burst exhibited a time profile that was well correlated with microwaves, El Leoncito 45, 90, and 212 GHz observations, Reuven Ramaty High-Energy Solar Spectroscopic Imager \citep[RHESSI;][]{2002SoPh..210....3L} and Fermi \citep{2009ApJ...697.1071A} high-energy X-rays, and white-light data \citep{2013ApJ...768..134K}. The 30 THz emission was spatially consistent with the white-light source, RHESSI HXR-rays (HXRs), extreme ultra-violet (EUV), and H$\alpha$ sources, exhibiting a peak flux of 12000 sfu (1 solar flux unit (sfu) = $10^{-22}\,$W$\,$m$^{-2}$Hz$^{-1}$). 

The largest 30 THz impulsive burst so far was observed from a GOES class X2 flare on 27 October 2014, exhibiting 35000 sfu \citep{2015JGRA..120.4155K}. The 30 THz peak was co-aligned in space with a pair of brightenings that were visible in both H$\alpha$ and white-light data, and well correlated in time to sub-THz peaks at 0.2 and 0.4 THz, RHESSI hard X-rays, and white light. 

These first discoveries suggest that 30 THz impulsive emission may arise lower in the denser chromosphere, and at the same site produce the other observed radiations. The 30 THz emission might be an extension of the synchrotron spectrum that produces the sub-THz emissions. However, the good correlation with hard X-rays and white-light emission may also favor thermal heating of the dense lower solar atmosphere by high-energy particle beams as a thermal back-warming process; see \cite{1970SoPh...15..176N,1989SoPh..124..303M,2006ApJ...641.1210X,2012ApJ...753L..26M}.

We report here the first 30 THz impulsive flare observed by the new S\~{a}o Paulo telescope \citep{2015SoPh..tmp..116K}. This event exhibits more complex spatial and temporal relationships between the 30 THz time profile and the microwave, H$\alpha$, EUV, soft X-ray (SXR), and metric$-$decimetric (m$-$dm) emissions than was seen in the two earlier events.

\section{Observations}

A GOES SXR flare of class M2 was reported by NOAA Space Weather Prediction Center\footnote{\url{http://www.swpc.noaa.gov/Data/goes.html}} on 1 August 2014 (14:43/14:48/14:57 UT) in active region 12130  (S09E35). This was the first impulsive burst observed by the new 30 THz telescope in Brazil \citep{2015SoPh..tmp..116K}. It was also observed at H$\alpha$ by the H-Alpha Solar Telescope of Argentina \citep[HASTA;][]{1999ESASP.448..469B} and at sub-mm wavelengths by Solar Submillimeter-wave Telescope \citep[SST;][]{2008SPIE.7012E..0LK}. EUV observations were obtained by the Atmospheric Imaging Assembly (AIA) on the Solar Dynamics Observatory \citep[SDO;][]{2012SoPh..275...17L}, and white-light observations were provided by the Helioseismic and Magnetic Imager \citep[HMI;][]{2012SoPh..275..207S} on SDO. Microwave observations were obtained by U.S. Air Force Radio Solar Telescope Network \citep[RSTN;][]{1979BAAS...11..311G}. Complementary metric-decimetric burst data were made available by RSTN, by Ondrejov Observatory in Czech Republic \citep{1993SoPh..147..203J}, and Nan\c{c}ay Observation Radiospectrographique pour Fedome et l'Etude des Eruptions Solaires (ORFEES) spectrometer in France \citep{2013URSI...K}.

The 30 THz observation and calibration methods are described in detail in \cite{2013ApJ...768..134K,2015JGRA..120.4155K,2015SoPh..tmp..116K}. The same procedures were used here. The 30 THz frames obtained at a cadence of five per second were integrated to 1 second sampling. The smoothed over 10 point curve has a maximum at 14:47:26 UT. The spatial resolution was of the order of 15 arcsec, set by the diffraction limit of the 15 cm aperture.

\begin{figure}[t!]
\includegraphics[width=0.56\textwidth,clip=]{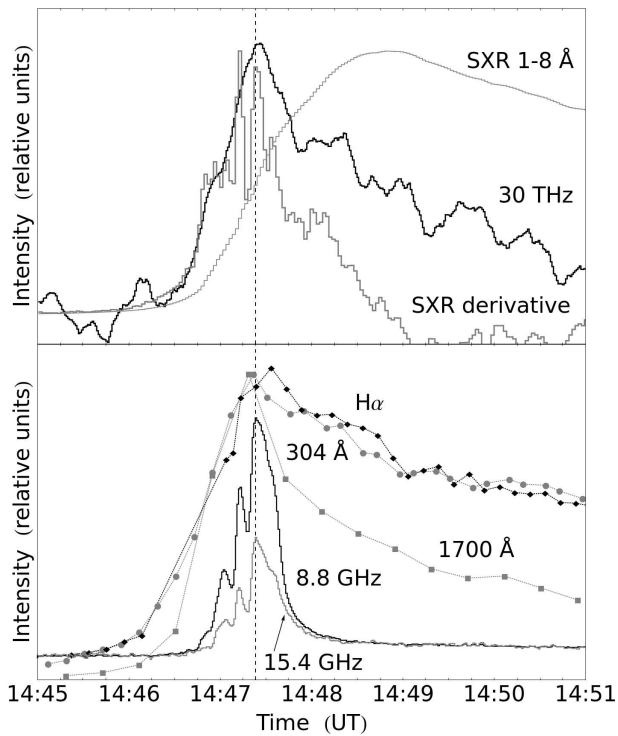}  
\caption[]{Time profiles with intensity in relative scales. Upper panel: 30 THz (black solid line), GOES 1$-$8~\AA{} (thin gray), and derivative (thick gray line). Lower panel: H$\alpha$ (black diamonds), SDO/EUV 304~\AA{}~(gray dots), 1700~\AA{}~continuum (gray squares), and RSTN higher frequency microwaves (solid black and gray line). The vertical dashed line at the microwave maximum (14:47:23 UT) is for a comparison purpose.}
\label{Fig1}
\end{figure}

The detailed time profiles at 30 THz and other wavelengths are shown in Fig.~\ref{Fig1}. The onset in the derivative of the GOES 1$-$8~\AA{} data matches the 30 THz onset time. This suggests a correlation with the rise time in HXRs, according to the Neupert effect \citep{1968ApJ...153L..59N}. However, the two peaks seen near the maximum in the SXR derivative have no correspondence to any features at 30 THz. The higher frequency radio time profiles shown in the bottom panel\footnote{The event was observed by both the Sagamore Hill and San Vito RSTN sites, and we found a difference in timing that required correction. We achieved this by comparing the RSTN 5 GHz data with the Ondrejov 5 GHz profile (not shown here). We advanced the RSTN timing by 5 seconds at Sagamore Hill and by 3
seconds at San Vito, with an uncertainty of order 1 second. The data shown result from merging the two RSTN datasets to fill in sampling gaps in both near the burst peak. Fluxes at the same frequency at the two RSTN observatories agreed to within 10\%. For more details, see Appendix.} exhibit three successive pulses 10 seconds apart during the rise to the impulsive peak. Peak fluxes were about 200 sfu at 15.4 GHz and 400 sfu at 8.8 GHz, in both cases at 14:47:23 UT. The microwave pulses are not prominent at optically thick frequencies below the spectral peak at around 8.8 GHz. The two strongest microwave pulses match the two prominent peaks in the GOES derivative, confirming that the SXR emission at this time results from heating by nonthermal electrons. 

We emphasize that there is a delay of order 1 minute of the microwave onset relative to the 30 THz onset. This suggests that the initial energy release did not produce significant nonthermal electrons (which radiate at microwave frequencies prolifically) in the corona.  The bottom panel also shows that the H$\alpha$ and EUV channels rise early in conjunction with the 30 THz rise time. The decay phase at 1700~\AA{} is similar in behavior to the 30 THz decay, while the H$\alpha$ and 304~\AA{} emissions decay at a slower rate.

The peak flux density at 30 THz is estimated to be 19000 sfu with an uncertainty of about $\pm$25\% (see the earlier references for a discussion of the calibration uncertainties). The event was not detected by SST at 0.2 or 0.4 THz with upper limits of about 10 sfu. Dynamic radio spectra and single-frequency time profiles at m$-$dm wavelengths are provided by Ondrejov, ORFEES, and RSTN. Ondrejov and ORFEES show type III reverse drift bursts near the peak of the 30 THz burst as well as several complex emissions in the corona whose description is beyond the scope of this RN. 

\begin{figure}[ht!]
\centerline{
\includegraphics[width=0.29\textwidth,clip=]{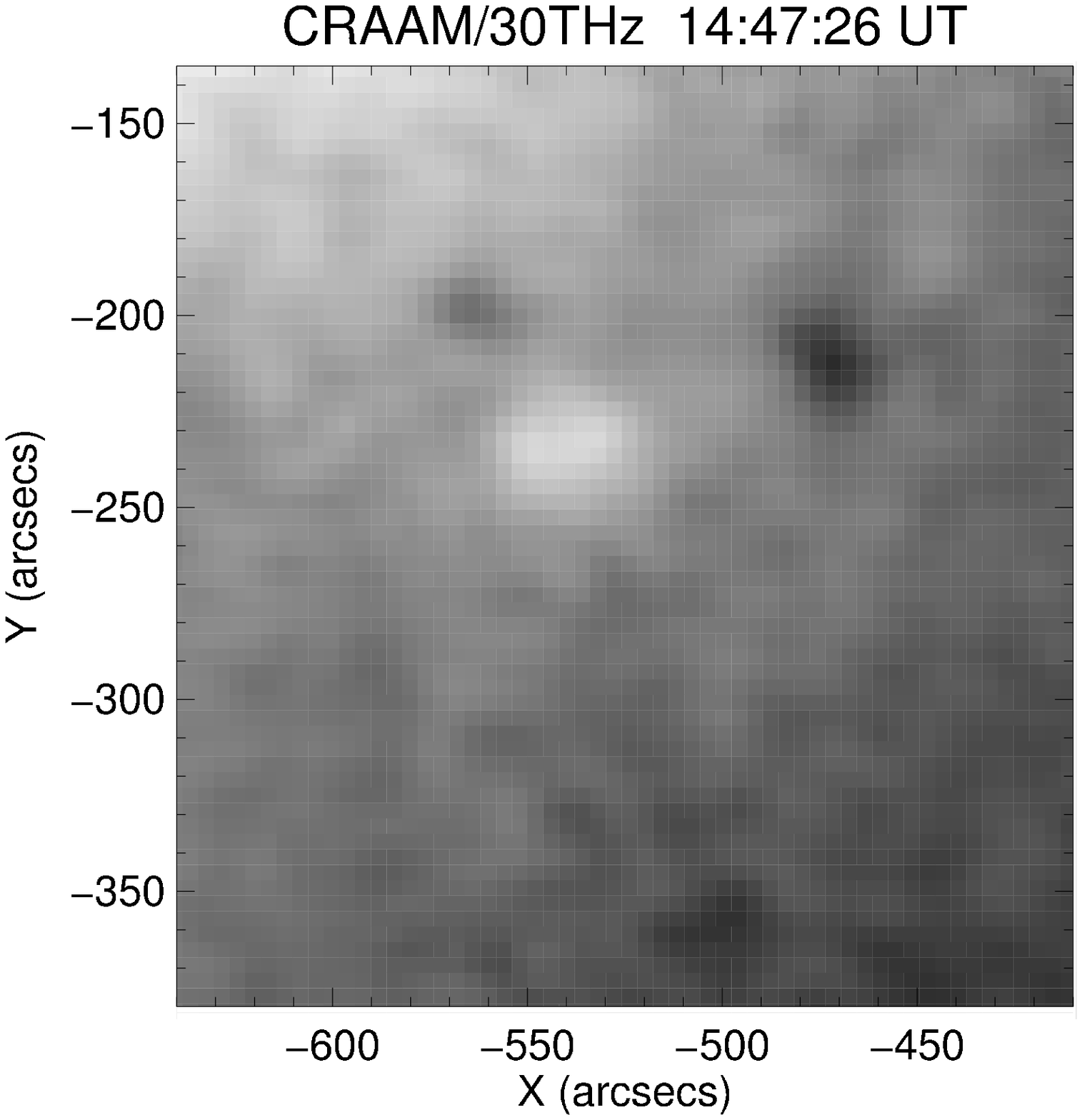}}
\vspace{0.01\textwidth}
\centerline{
\includegraphics[width=0.29\textwidth,clip=]{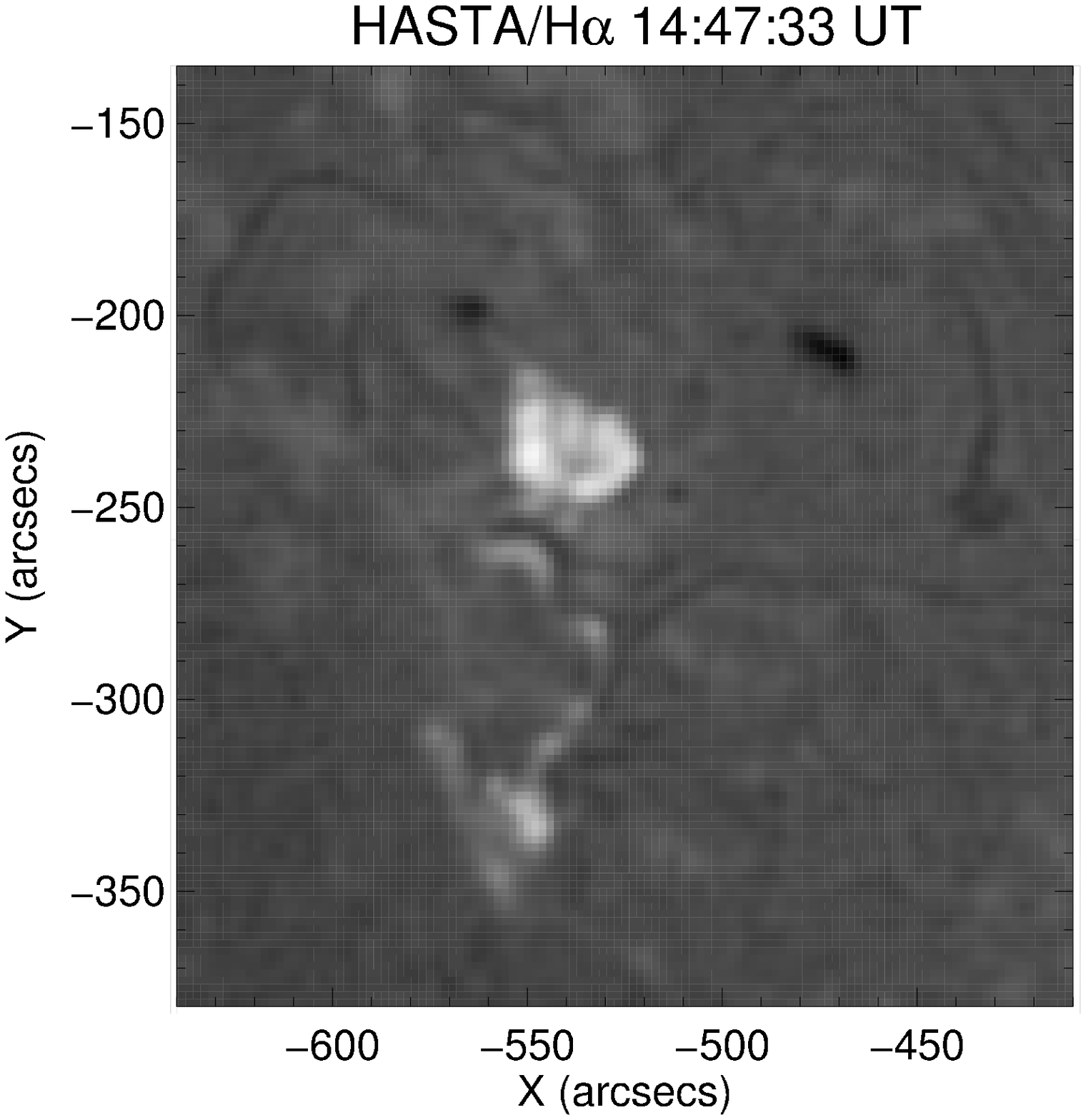}}
\vspace{0.01\textwidth}
\centerline{
\includegraphics[width=0.29\textwidth,clip=]{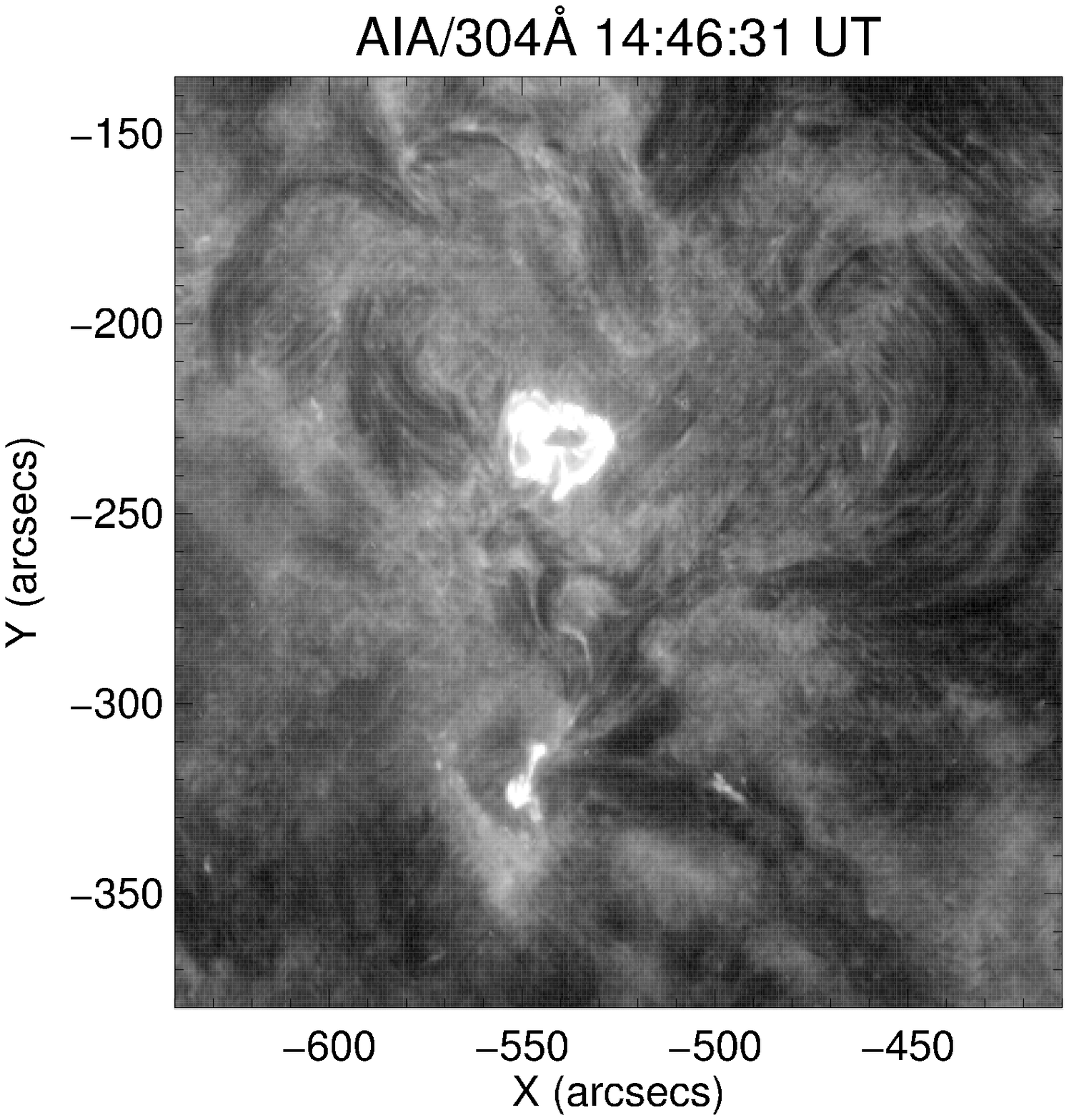}}
\vspace{0.01\textwidth}
\centerline{
\includegraphics[width=0.29\textwidth,clip=]{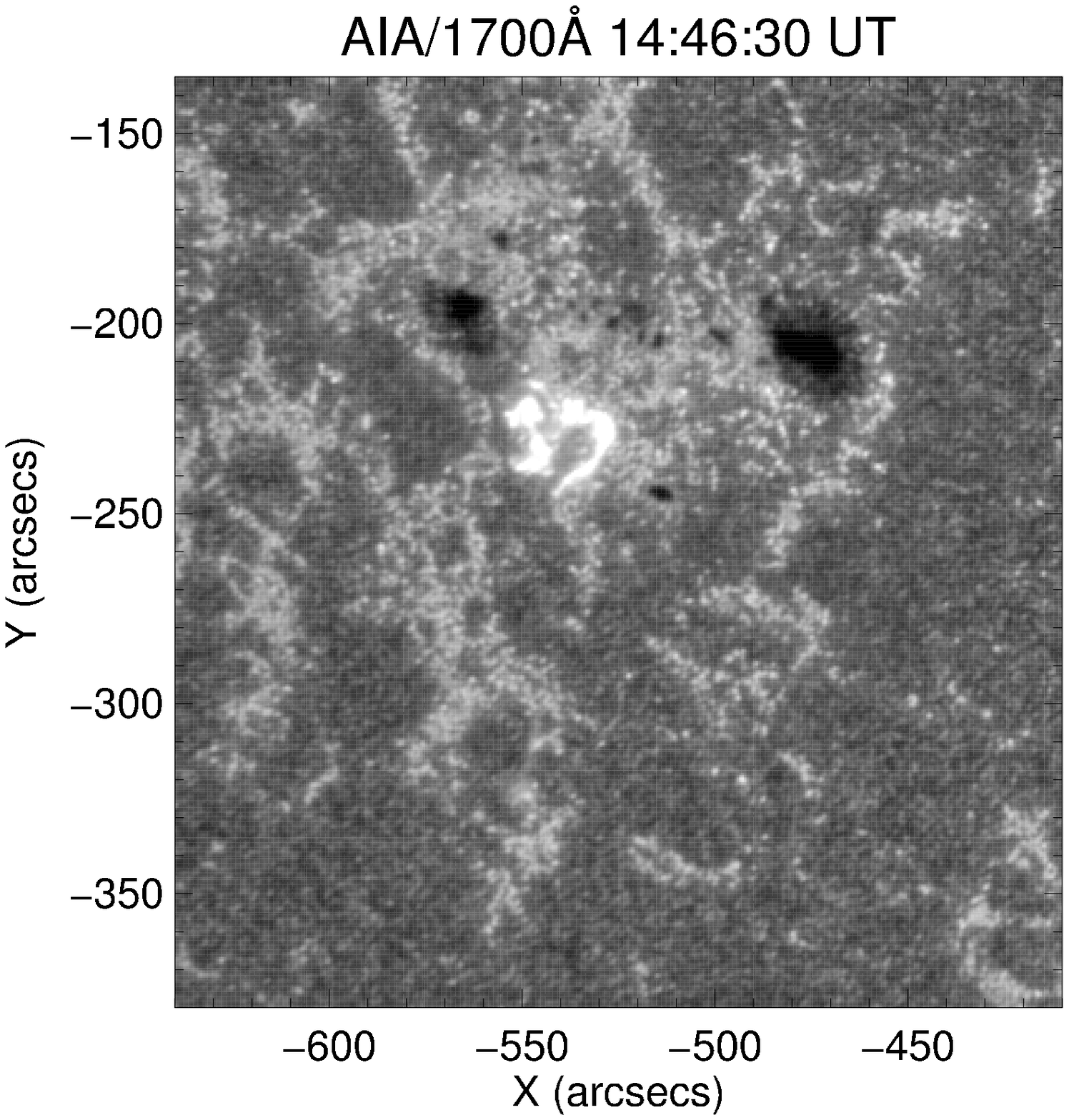}}
\vspace{-0.01\textwidth}
\caption[]{Flare images at about the maximum emission time: 30 THz from Centro de R\'{a}dio Astronomia e Astrof\'{i}sica Mackenzie (CRAAM), H$\alpha$ from HASTA, EUV 304~\AA,{} and 1700~\AA{} from SDO/AIA.}
\label{Fig2}
\end{figure}

The 30 THz image near the time of maximum emission is shown in Fig.~\ref{Fig2}, and exhibits a spatial co-alignment with the flare image at about the same time obtained by HASTA in H$\alpha$ and by AIA/SDO at 304~\AA{} (transition region) and 1700~\AA{} (chromosphere) wavelengths. The fine-scale structure visible in the HASTA/H$\alpha$ and EUV images is too small scale to be resolved by the 30 THz system. There was no detectable white-light emission in the HMI images.

\section{Discussion and conclusions}
\label{Sect_Disc}

The rise phase of the 30 THz emission seems to match those in H$\alpha$, 304~\AA,{}  1700~\AA,{} and the SXR derivative (the HXR proxy) rise phase. By contrast, the nonthermal microwave signature does not show such an early rise. The 30 THz decay phase appears to be very similar to that of the 1700~\AA{} continuum. The H$\alpha$ and 304~\AA{} emissions, however, are similar to each other, but decay at a slower rate following the peak than the 30 THz emission. These features come from very different layers of the solar atmosphere: 304~\AA\ from the transition region and corona, H$\alpha$ from the upper chromosphere, and 1700~\AA\ from the chromosphere. Different time constants in different layers of the atmosphere can explain some of the differences in time profiles. Thus, optically thin coronal sources can radiate energy away rapidly, while sources deep in the chromosphere and features such as the SXR emission, which are believed to be related to the time integral of the nonthermal heating, may take longer to dissipate the energy deposited there. This effect  tends to smooth out fine time structures in the energy release. This may explain why the pulses seen in the rise of the high-frequency microwave emissions are not manifested at other wavelengths. The absence of such structures in the 30 THz emission seems to argue against a direct nonthermal source for the mid-IR emission in this event. Unfortunately, we have no imaging in either microwaves or X-rays that might test this idea: if the 30 THz emission is nonthermal, we expect that it is consistent spatially (either coincident, or in the case of HXR, in the corona above chromospheric footpoints) with the other nonthermal diagnostics.

The intense 30 THz peak flux (19000 sfu) is nearly 50 times larger than the microwave flux, and even larger than the upper limits at 0.2 and 0.4 THz. If we extrapolate the 30 THz flux down to 0.4 THz with a frequency$-$squared spectrum (typical of an optically thick isothermal source, and flatter than the 2.5$-$3.0 spectral index typical of optically thick nonthermal gyrosynchrotron emission), we would predict that there would only be 3 sfu. This argues against the possibility that a single mechanism can explain the combined spectrum from microwave frequencies up to 30 THz, and favors the presence of (at least) two components in this spectral range, as found for the two other 30 THz bursts analyzed previously \citep{2013ApJ...768..134K,2015JGRA..120.4155K}. 

The coincidence of the rise phases of the 30 THz, EUV, and H$\alpha$ emission and the spatial coincidence of the brightenings suggest that they all originate from the same location somewhere in the lower chromosphere. The similarity of the 30 THz rise and fall time profile  with that of the 1700~\AA{} continuum suggests that the origin of both emissions might be close to the loop footpoints. If the 30 THz emission originates near the temperature minimum, as argued by \citet{2013ApJ...768..134K}, then we would also expect white-light emission, and we have to attribute the lack of detected white-light emission to the sensitivity limitations of HMI. The 304~\AA{} He II and H$\alpha$ lines are known to be produced higher at the upper chromosphere$-$lower transition region, which possibly explais the observed slower decay rate compared to the lower-atmosphere signatures.

While the coincidence in timing of the main peak at most of the wavelengths available clearly suggests a common origin, there are sufficient differences in the timing of the onset of emission to present a puzzle. The presence of nonthermal electrons at the onset of 30 THz emission is suggested by the simultaneous onset of HXRs (inferred from the SXR derivative via the Neupert effect), but the lack of microwave emission until later suggests that there were no electrons in the corona with energies much above a few tens of keV. While the 30 THz emission near the peak is consistent with the standard picture of sudden plasma heating in response to high-energy nonthermal electron bombardment on the dense lower chromosphere regions \citep[as for white light and 1700 ~\AA\ emission, e.g.,][]{1970SoPh...15..176N,1989SoPh..124..303M,2006ApJ...641.1210X,2007ChJAA...7..721W,2012ApJ...753L..26M,2013ApJ...774...14Q}, the early rise of 30 THz and 1700 \AA\ emission before the microwaves does not seem consistent with this mechanism because electrons with energies of just a few tens of keV cannot penetrate to the depth required.

\begin{acknowledgements}
We thank R.~V.~Souza for help with the 30 THz data analysis. This work was partially supported by Brazilian agencies FAPESP (Proc.~2013/24155$-$3), CNPq, and Mackpesquisa, and the U.S. agency AFOSR. DPC acknowledges support from CAPES-PROSUP. MK acknowledges support from Grant P209/12/0103 (GA\v{C}R) and the EU FP7 project No. 606862 F-CHROMA. RM acknowledges CNPq project No.~23/2009.
\end{acknowledgements}


\bibpunct{(}{)}{;}{a}{}{,}

\bibliographystyle{aa}

\bibliography{bibref}

\IfFileExists{\jobname.bbl}{} {\typeout{}
\typeout{****************************************************}
\typeout{****************************************************}
\typeout{** Please run "bibtex \jobname" to obtain} \typeout{**
the bibliography and then re-run LaTeX} \typeout{** twice to fix
the references !}
\typeout{****************************************************}
\typeout{****************************************************}
\typeout{}}

\begin{appendix}

\section{Time correction of RSTN data}

For this event, radio data was available from both the Sagamore Hill and San Vito RSTN sites. The time shift found in the RSTN data was corrected by comparing the intensity$-$time profiles at RSTN with those observed at Ondrejov observatory. Ondrejov spectrograph data is available at 0.1 second resolution and the time is taken from the high-precision signal of an atomic clock. To determine the timing errors, we carried out a cross-correlation of the light curve deduced from the Ondrejov dynamic spectrum, summed from 4.8 to 5 GHz and smoothed to 1 second, with the 5 GHz light curve from the RSTN observatories. Ondrejov and RSTN light curves align, after advancing the RSTN timing by 5 seconds at Sagamore Hill and by 3 seconds at San Vito, with an uncertainty of order 1 second. After the applied time shift, the correlation between the RSTN light curve (Sagamore Hill and San Vito) with the Ondrejov data was found to be over 0.99. 

Both sets of RSTN data have short gaps (of a few seconds) at different times during the period of interest. This was accounted for, first, by shifting the datasets by the measured amounts to the correct time and, then, by averaging them together. The radio data shown in this work results from combining the two RSTN datasets to fill in sampling gaps near the burst peak. The difference in the fluxes recorded at Sagamore Hill and San Vito stations at the same frequency is less than 10\%. 

The same procedure was applied at the 2.7 GHz frequency level. The absence of sharp features in the light curves, however, makes this frequency less suitable for determining timing offsets and, thus, the 5 GHz is selected instead.
\end{appendix}

\end{document}